\documentclass[preprint,12pt]{elsarticle}
\usepackage[utf8x]{inputenc}

\usepackage{amsmath,amssymb}
\usepackage{xcolor}
\usepackage{float} 

\journal{Physica A}

\begin{document}

\begin{frontmatter}

\title{Entropic properties of $D$-dimensional Rydberg systems}

\author[label1]{I. V. Toranzo}
\address[label1]{Departamento de F\'{\i}sica At\'{o}mica, Molecular y Nuclear, Universidad de Granada, Granada 18071, Spain\\ Instituto Carlos I de F\'{\i}sica Te\'orica y Computacional, Universidad de Granada, Granada 18071, Spain}

\author[label1]{D. Puertas-Centeno}

\author[label1]{J. S. Dehesa}
\ead{dehesa@ugr.es}

\begin{abstract}
The fundamental information-theoretic measures (the Rényi $R_{p}[\rho]$ and Tsallis $T_{p}[\rho]$ entropies, $p>0$) of the highly-excited (Rydberg) quantum states of the $D$-dimensional ($D>1$) hydrogenic systems, which include the Shannon entropy ($p \to 1$) and the disequilibrium ($p = 2$), are analytically determined by use of the strong asymptotics of the Laguerre orthogonal polynomials which control the wavefunctions of these states. We first realize that these quantities are derived from the entropic moments of the quantum-mechanical probability $\rho(\vec{r})$ densities associated to the Rydberg hydrogenic wavefunctions $\Psi_{n,l,\{\mu\}}(\vec{r})$, which are closely connected to the $\mathfrak{L}_{p}$-norms of the associated Laguerre polynomials. Then, we determine the ($n\to\infty$)-asymptotics of these norms in terms of the basic parameters of our system (the dimensionality $D$, the nuclear charge and the hyperquantum numbers $(n,l,\{\mu\}$) of the state) by use of recent techniques of approximation theory. Finally, these three entropic quantities are analytically and numerically discussed in terms of the basic parameters of the system for various particular states. 
\end{abstract}



\end{frontmatter}


\section{Introduction}\label{s1}

Rydberg systems are ballooned-up atoms which can be made by exciting the outermost electron in certain elements, so that all the inner electrons can lumped together and regarded, along with the atom's nucleus, as a unified core, with the lone remaining electron lying outside \cite{lundee,gallagher}. Thus, they are as if the atom were a  hydrogenic system, a heavy version of hydrogen. Due to their extraordinary properties (high magnetic susceptibility, \textcolor{red}{relatively} long lifetime, high kinetic energy,…) these systems, where the outermost electrons are highly excited but not ionized, have been used in multiple scientific areas ranging from plasmas and diamagnetism to astrophysics, quantum chaos and strongly interacting systems. Recently it has been argued that they might be just the basic elements for processing quantum information (see e.g., \cite{shiell,saffman}). Indeed, these outsized atoms can be sustained for a long time in a quantum superposition condition (what is very convenient for creating qubits) and they can interact strongly with other such atoms; this property makes them very useful for devising the kind of logic gates needed to process information.\\ 
 
The $D$-dimensional hydrogenic system (i.e. an electron or a negatively-charged particle moving around a nucleus or a positively-charged core which electromagnetically binds it in its orbit), with $D>1$, is the main prototype to model the behavior of most multidimensional quantum many-body systems with standard ($D = 3$) and non-standard ($D\neq 3$) dimensionalities \cite{witten,herschbach_93, burgbacher:jmp99, aquilanti:cp97, aquilanti_96, andrew:ajp90}. It embraces a large variety of three-dimensional physical systems (e.g., hydrogenic atoms and ions, exotic atoms, antimatter atoms, Rydberg atoms,…) and a number of nanotechnological objects which have been shown to be very useful in semiconductor nanostructures (e.g., quantum wells, wires and dots) \cite{harrison_05, li:pla07} and quantum computation (e.g., qubits) \cite{nieto:pra00, dykman:prb03}. Moreover, it plays a crucial role for the interpretation of numerous phenomena of quantum cosmology \cite{amelinocamelia_05} and quantum field theory \cite{witten, itzykson_06,dong}. As well, the $D$-dimensional hydrogenic wavefunctions have been used as complete orthonormal sets for many-body atomic and molecular problems \cite{aquilanti:aqc01, aquilanti:irpc01,coletti} in both position and momentum spaces. Finally, the existence of non-standard hydrogenic systems has been proved for $D<3$ \cite{li:pla07,caruso} and suggested for $D>3$ \cite{burgbacher:jmp99}.\\

 The multidimensional extension of Rydberg hydrogenic states (i.e. states where the electron has a large principal quantum number $n$, so being highly excited), with standard and non-standard dimensionalities, has been investigated (see section 5 of \cite{dehesa_2010}, and \cite{tarasov,lopez_2009,aptekarev_2010,dehesa_sen12}) up until now by means of the following spreading measures: central moments, variances, logarithmic expectation values, Shannon entropy and Fisher information. These measures were found to be expressed in terms of the principal and orbital hyperquantum numbers and the space dimensionality $D$. In this work we go much beyond this study by calculating the Rényi \cite{renyi1} and Tsallis \cite{tsallis} entropies (also called by \textit{information generating functionals} \cite{golomb}) of the Rydberg states defined by
\begin{eqnarray}
\label{eq:renentrop}
R_{p}[\rho] &=&  \frac{1}{1-p}\ln W_{p}[\rho]; \quad 0<p<\infty,\\
\label{eq:tsalentrop}
T_{p}[\rho] &=& \frac{1}{p-1}(1-W_{p}[\rho]); \quad 0<p<\infty,
\end{eqnarray}
where the symbol $W_{p}[\rho]$ denotes the entropic moments of $\rho(\vec{r})$ defined as 
\begin{equation}
\label{eq:entropmom}
W_{p}[\rho] = \int_{\mathbb{R}^D} [\rho(\vec{r})]^{p}\, d\vec{r} =\| \rho\|_p^p;\quad p > 0. 
\end{equation}
The symbol $\|\cdot\|_p$     denotes    the    $\mathfrak{L}_{p}$-norm   for    functions:
$\|\Phi\|_p=\left(\int_{\mathbb{R}^D} |\Phi(\vec{r})|^p d\vec{r}\right)^{1/p}$. 
Note that both Rényi and Tsallis measures include the Shannon entropy, $S[\rho] = \lim_{p\rightarrow 1} R_{p}[\rho] = \lim_{p\rightarrow 1} T_{p}[\rho]$, and the disequilibrium, $\langle\rho\rangle = \exp(R_{2}[\rho])$, as two important particular cases. Moreover, they are interconnected as indicated later on. Their properties have been recently reviewed \cite{dehesa_sen12,bialynicki3,jizba}; see also \cite{aczel,dehesa_88,dehesa_89,romera_01,leonenko,guerrero}. Moreover, the R\'enyi entropies and their associated uncertainty relations have been widely used to investigate a great deal of quantum-mechanical properties and phenomena of physical systems and processes \cite{bialynicki2,dehesa_sen12,bialynicki3}, ranging from the quantum-classical correspondence \cite{sanchezmoreno} and quantum entanglement \cite{bovino} to pattern formation and Brown processes \cite{cybulski1,cybulski2}, quantum phase transition \cite{calixto} and disordered systems \cite{varga}.

The structure of this work is the following. First, in Section II, we give the wavefunctions of the stationary $D$-dimensional hydrogenic states in position space and their squares, the quantum probability densities $\rho\left(\vec{r} \right)$. Then we define the entropic moments and the Rényi entropy of this density, and we show that for the very excited states the calculation of the latter quantity essentially converts into the determination of the asymptotics of the $\mathfrak{L}_{p}$-norm of the Laguerre polynomials which control the states' wavefunctions. In Section III we use a powerful technique of approximation theory recently developed by Aptekarev et al \cite{aptekarev_2012,aptekarev_2010,aptekarev_2016} to determine these Laguerre norms in terms of $D$ and the hyperquantum numbers of the system. In Section IV the Shannon, Rényi and Tsallis entropies are studied both analytically and numerically for the $D$-dimensional hydrogenic states by means of $D$, the hyperquantum numbers and the nuclear charge $Z$ of the system. Finally, some concluding remarks are given.

\section{The $D$-dimensional Rydberg problem: entropic formulation}

In this section we briefly describe the quantum probability density of the stationary states of the $D$-dimensional hydrogenic system in position space. Then, we pose the determination of the entropic moments and the Rényi entropies of this density in the most appropriate mathematical manner for our purposes. Atomic units will be used throughout.\\
The time-independent Schr\"{o}dinger equation of a $D$-dimensional ($D \geqslant 1$) hydrogenic system (i.e., an electron moving under the action of the $D$-dimensional Coulomb potential $\displaystyle{V(\vec{r})=-\frac{Z}{r}}$) is given by
\begin{equation}\label{eqI_cap1:ec_schrodinger}
\left( -\frac{1}{2} \vec{\nabla}^{2}_{D} - \frac{Z}{r}
\right) \Psi \left( \vec{r} \right) = E \Psi \left(\vec{r} \right),
\end{equation}
where $\vec{\nabla}_{D}$ denotes the $D$-dimensional gradient operator, $Z$ is the nuclear charge, and the  electronic position vector $\vec{r}  =  (x_1 ,  \ldots  , x_D)$ in hyperspherical units  is  given as $(r,\theta_1,\theta_2,\ldots,\theta_{D-1})      \equiv
(r,\Omega_{D-1})$, $\Omega_{D-1}\in S^{D-1}$, where $r \equiv |\vec{r}| = \sqrt{\sum_{i=1}^D x_i^2}
\in [0  \: ;  \: +\infty)$  and $x_i =  r \left(\prod_{k=1}^{i-1}  \sin \theta_k
\right) \cos \theta_i$ for $1 \le i \le D$
and with $\theta_i \in [0 \: ; \: \pi), i < D-1$, $\theta_{D-1} \equiv \phi \in [0 \: ; \: 2
\pi)$. It is assumed that the nucleus is located at the origin and, by  convention, $\theta_D =  0$ and the  empty product is the  unity. . \\
It is known \cite{nieto,yanez_1994,dehesa_2010,dong} that the energies belonging to the discrete spectrum are given by  
\begin{equation} \label{eqI_cap1:energia}
E= -\frac{Z^2}{2\eta^2},\hspace{0.5cm} \hspace{0.5cm} \eta=n+\frac{D-3}{2}; \hspace{5mm} n=1,2,3,...,
\end{equation}
and the associated eigenfunction can be expressed as
\begin{equation}\label{eqI_cap1:FunOnda_P}
\Psi_{\eta,l, \left\lbrace \mu \right\rbrace }(\vec{r})=\mathcal{R}_{\eta,l}(r)\,\, {\cal{Y}}_{l,\{\mu\}}(\Omega_{D-1}).
\end{equation}
Then, the quantum probability density of a $D$-dimensional hydrogenic stationary state $(n,l,\{\mu\})$ is given by the squared modulus of the position eigenfunction as 
\begin{equation}
\label{eq:denspos}
\rho_{n,l,\{\mu\}}(\vec{r}) = \rho_{n,l}(\tilde{r})\,\, |\mathcal{Y}_{l,\{\mu\}}(\Omega_{D-1})|^{2},
\end{equation}
where the radial part of the density is the univariate function
\begin{eqnarray}
\label{eq:radensity}
\rho_{n,l}(\tilde{r}) &=& [\mathcal{R}_{n,l}(r)]^2 = \frac{\lambda^{-d}}{2 \eta} \,\, \frac{\omega_{2L+1}(\tilde{r})}{\tilde{r}^{d-2}}\,\,[{\widehat{\mathcal{L}}}_{\eta-L-1}^{(2L+1)}(\tilde{r})]^{2}
\end{eqnarray}
with $L$, defined as the ``grand orbital angular momentum quantum number'', and the adimensional parameter $\tilde{r}$ given by
\begin{align} \label{eqI_cap1:Lyr}
L&=l+\frac{D-3}{2}, \hspace{0.5cm} l=0, 1, 2, \ldots \\ \label{rtilde}
\tilde{r}&=\frac{r}{\lambda},\hspace{0.5cm} 
\hspace{0.5cm}\lambda=\frac{\eta}{2Z}.
\end{align}
The symbols $\mathcal{L}_{n}^{(\alpha)}(x)$ and $\widehat{\mathcal{L}}_{n}^{(\alpha)}(x)$ denote the orthogonal and orthonormal, respectively, Laguerre polynomials with respect to the weight $\omega_\alpha(x)=x^{\alpha} e^{-x}$ on the interval $\left[0,\infty \right) $, so that
\begin{equation}\label{eqI_cap1:laguerre_orto_ortogo}
{\widehat{\mathcal{L}}}^{(\alpha)}_{m}(x)=  \left( \frac{m!}{\Gamma(m+\alpha+1)}\right)^{1/2}  {\mathcal{L}}^{(\alpha)}_{m}(x),
\end{equation}
and finally
\begin{equation}
K_{\eta,L} = \lambda^{-\frac{D}{2}}\left\{\frac{(\eta-L-1)!}{2\eta(\eta+L)!}\right\}^{\frac{1}{2}}=\left\{\left(\frac{2Z}{n+\frac{D-3}{2}}\right)^{D}\frac{(n-l-1)!}{2\left(n+\frac{D-3}{2}\right)(n+l+D-3)!}  \right\}^{\frac{1}{2}} \equiv K_{n,l} 
\label{eq:4}
\end{equation}
represents the normalization constant which ensures that $\int \left| \Psi_{\eta,l, \left\lbrace \mu \right\rbrace }(\vec{r}) \right|^2 d\vec{r} =1$. 
The angular eigenfunctions are the hyperspherical harmonics, $\mathcal{Y}_{l,\{\mu\}}(\Omega_{D-1})$, defined as
\begin{equation}
\label{eq:hyperspherarm}
\mathcal{Y}_{l,\{\mu\}}(\Omega_{D-1}) = \mathcal{N}_{l,\{\mu\}}e^{im\phi}\nonumber\times \prod_{j=1}^{D-2}\mathcal{C}^{(\alpha_{j}+\mu_{j+1})}_{\mu_{j}-\mu_{j+1}}(\cos\theta_{j})(\sin\theta_{j})^{\mu_{j+1}}
\end{equation}
with the normalization constant
\begin{equation}
\label{eq:normhypersphar}
\mathcal{N}_{l,\{\mu\}}^{2} = \frac{1}{2\pi}\times\nonumber\\
\prod_{j=1}^{D-2} \frac{(\alpha_{j}+\mu_{j})(\mu_{j}-\mu_{j+1})![\Gamma(\alpha_{j}+\mu_{j+1})]^{2}}{\pi \, 2^{1-2\alpha_{j}-2\mu_{j+1}}\Gamma(2\alpha_{j}+\mu_{j}+\mu_{j+1})},\nonumber\\
\end{equation}
where the symbol $\mathcal{C}^{\lambda}_{n}(t)$ denotes the Gegenbauer polynomial of degree $n$ and parameter $\lambda$.\\

Now we can calculate any spreading measure of the $D$-dimensional hydrogenic system beyond the known the variance and the ordinary moments (or radial expectation values) of its density, which are already known \cite{dehesa_2010}. The most relevant spreading quantities are the entropic moments $W_{p}[\rho_{n,l,\{\mu\}}]$, because they characterize the density and moreover we can obtain from them the main entropic measures of the system such as the Rényi, Shannon and Tsallis entropies. They are given as
\begin{eqnarray}
\label{eq:entropmom2}
W_{p}[\rho_{n,l,\{\mu\}}] &=& \int_{\mathbb{R}^D} [\rho_{n,l,\{\mu\}}(\vec{r})]^{p}\, d\,\vec{r}\nonumber\\ &=& \int\limits_{0}^{\infty}[\rho_{n,l}(r)]^{p}\,r^{D-1}\,dr\times \Lambda_{l,\{\mu\}}(\Omega_{D-1}),
\end{eqnarray}
where we have used that the volume element is
\[
d\vec{r} =r^{D-1}dr\,d\Omega_{D-1} , \quad d\Omega_{D-1} = \left(\prod_{j=1}^{D-2}\sin^{2\alpha_{j}}\theta_{j}\right)d\phi,
\]
(with $2\alpha_{j}= D-j-1$) and the angular part is given by
\begin{equation}
\label{eq:angpart}
\Lambda_{l,\{\mu\}}(\Omega_{D-1}) = \int_{S^{D-1}}|\mathcal{Y}_{l,\{\mu\}}(\Omega_{D-1})|^{2p}\, d\Omega_{D-1}.
\end{equation}
Then, from Eqs. (\ref{eq:renentrop}) and (\ref{eq:entropmom2}) we can obtain the Rényi entropies of the $D$-dimensional hydrogenic state $(n,l,\{\mu\})$ as follows
\begin{equation}
\label{eq:renyihyd1}
R_{p}[\rho_{n,l,\{\mu\}}] = R_{p}[\rho_{n,l}]+R_{p}[\mathcal{Y}_{l,\{\mu\}}],
\end{equation}
where $R_{p}[\rho_{n,l}]$ denotes the radial part
\begin{equation}
\label{eq:renyihyd2}
R_{p}[\rho_{n,l}] = \frac{1}{1-p}\ln \int_{0}^{\infty} [\rho_{n,l}]^{p} r^{D-1}\, dr,
\end{equation}
and  $R_{p}[\mathcal{Y}_{l,\{\mu\}}]$ denotes the angular part
\begin{equation}
\label{eq:renyihyd3}
R_{p}[\mathcal{Y}_{l,\{\mu\}}] = \frac{1}{1-p}\ln \Lambda_{l,\{\mu\}}(\Omega_{D-1}).
\end{equation}
In this work we are interested in the entropic properties of the high extreme region of the system, embracing the highly and very highly excited (Rydberg) states (recently shown to be experimentally accessible \cite{lundee}) where these properties are most difficult to compute because they possess large and very large values of $n$. Since the dependence of both the entropic moments and the Rényi entropies on the principal hyperquantum number $n$ is concentrated in their radial parts according to Eqs. (\ref{eq:entropmom2})-(\ref{eq:renyihyd3}), the computation of $R_{p}[\rho_{n,l,\{\mu\}}]$ for the Rydberg states of $D$-dimensional hydrogenics systems practically reduces to determine the value of the radial Rényi entropy, $R_{p}[\rho_{n,l}]$, in the limiting case $n\rightarrow \infty$. Moreover, by taking into account the expression (\ref{eq:radensity}) of the radial density and Eqs. (\ref{eq:entropmom2}) and (\ref{eq:renyihyd2}), this problem converts into the study of the asymptotics ($n\to\infty$) of the $\mathfrak{L}_{p}$-norm of the Laguerre polynomials 
\begin{equation}\label{eq:c1.2}
N_{n}(\alpha,p,\beta)=\int\limits_{0}^{\infty}\left(\left[\widehat{\mathcal{L}}_{n}^{(\alpha)}(x)\right]^{2}\,w_{\alpha}(x)\right)^{p}\,x^{\beta}\,dx\;,\quad \alpha >-1\, \, , p>0\,\, , \beta +p\alpha >-1.
\end{equation}
Indeed, from Eq. (\ref{eq:renyihyd2}) one has that the radial Rényi entropy can be expressed as 
\begin{equation}
\label{eq:renyihyd4}
R_{p}[\rho_{n,l}] = \frac{1}{1-p}\ln\left[\frac{\eta^{D(1-p)-p}}{2^{D(1-p)+p}Z^{D(1-p)}}N_{n,l}(D,p) \right],
\end{equation}
where the norm $N_{n,l}(\alpha,p,\beta) \equiv N_{n,l}(D,p)$ is given by
\begin{equation}\label{eq:c1.21}
N_{n,l}(D,p)=\int\limits_{0}^{\infty}\left(\left[\widehat{\mathcal{L}}_{n-l-1}^{(\alpha)}(x)\right]^{2}\,w_{\alpha}(x)\right)^{p}\,x^{\beta}\,dx,
\end{equation}
with 
\begin{equation}
\label{eq:condition}
\alpha=2L+1=2l+D-2\,,\;l=0,1,2,\ldots,n-1,\, p>0\,\, \text{and}\,\, \beta=(2-D)p+D-1
\end{equation}
We note that (\ref{eq:condition}) guarantees the convergence of
integral (\ref{eq:c1.21}); i.e. the condition $\beta+p\alpha= 2lp+D-1 > -1$ is always satisfied for physically meaningful values of the parameters.\\
 
\section{Laguerre $\mathfrak{L}_{p}$-norms and radial entropies: Asymptotics }

Let us now study the asymptotics at large $n$ of the Laguerre hydrogenic norms $N_{n,l}(D,p)$ given by Eq. (\ref{eq:c1.21}) in terms of all possible values of the involved parameters $(D,p)$. It controls the asymptotic values of the radial Rényi entropy $R_{p}[\rho_{n,l}]$ given by Eq. (\ref{eq:renyihyd4}) and, because of Eq. (\ref{eq:renyihyd1}), the total Rényi entropy of the Rydberg $D$-dimensional hydrogenic states. Since the exact evaluation of these norm-like functionals is a very difficult task, not yet solved, we will tackle this problem by means of the determination of the asymptotical behavior (i.e., at large $n$) of the general functional $N_{n}(\alpha,p,\beta)$ by extensive use of the strong asymptotics of Laguerre polynomials. The results obtained are contained in the following theorem.\\

\textbf{Theorem.}
The asymptotics ($n\to\infty$) of the Laguerre hydrogenic functionals $N_{n,l}(D,p)$ defined by Eq. (\ref{eq:c1.21}) are given by the following expressions for all possible values of $D$ and $p>0$:
\begin{enumerate}
\item If $\beta > 0$, there are two subcases:
	\begin{enumerate}
	\item If $D>2$, then
		\begin{equation}\label{13}
			N_{n,l}(D, p)= C(\beta,p)\,(2(n-l-1))^{1+\beta-p}\,(1+\bar{\bar{o}}(1)),\; \text{for}\,\, p\in \left(0,\frac{D-1}{D-2}\right)
		\end{equation}
	\item If $D=2$ (so, $\beta=1$), then
	\begin{equation}\label{14}
		N_{n,l}(D, p)=\left\{
		\begin{array}{ll}
		C(1,p)\,(2(n-l-1))^{2-p}\,(1+\bar{\bar{o}}(1))\;, & p\in (0,2)
		\\
		\displaystyle\frac{\ln (n-l-1)+\underline{\underline{O}}(1)}{\pi^{2}}\;, & p=2(*)
		\\
		\displaystyle\frac{C_{A}(p)}{\pi^p}\,(4(n-l-1))^{\frac{2}{3}(2-p)}(1+\bar{\bar{o}}(1))\, & p\in(2,5)
		\\
		\left(\displaystyle\frac{C_{A}(p)}{\pi^p 4^{2}}+C_{B}(\alpha,1,p)\right)\,(n-l-1)^{-2},
		& p=5
		\\
		C_{B}(\alpha,1,p)\,(n-l-1)^{-2}\;, & p\in(5,\infty) .
		\end{array}
		\right.
		\end{equation}
(*) Cosine-Airy regime
	\end{enumerate}
\item If $\beta = 0$ (so, $D\neq 2$ and $p=\frac{D-1}{D-2}$), then
\begin{equation}\label{9}
N_{n,l}(D, p)=\left\{
\begin{array}{ll}
C(0,p)\,(2(n-l-1))^{(1-p)}\,(1+\bar{\bar{o}}(1))\;, & p=\frac{D-1}{D-2}
\\
\displaystyle\frac{\ln (n-l-1)+\underline{\underline{O}}(1)}{\pi^{2}(n-l-1)}\;, & p=2,\, (D=3)(*)
\end{array}
\right.
\end{equation}
(*)Cosine-Airy regime
\item If $\beta < 0$ (so, either $p<\frac{D-1}{D-2}$ and $D<2$ or $p>\frac{D-1}{D-2}$ and $D>2$), then
\begin{equation}\label{8}
N_{n,l}(D, p)=\left\{
\begin{array}{ll}
C(\beta,p)\,(2(n-l-1))^{1+\beta-p}\,(1+\bar{\bar{o}}(1)),\quad & p\in \left( \frac{D-1}{D-2}, \frac{2 D}{2 D-3}\right)
\\
\displaystyle\frac{2\,\Gamma(p+1/2)\,\,\,(\ln n+\underline{\underline{O}}(1))}{\pi^{p+1/2}\,\Gamma(p+1)\,(4(n-l-1))^{1+\beta}}\,,\quad & p=2+2\,\beta = \frac{2 D}{2 D-3}(*),
\\
C_{B}(\alpha,\beta,p)\,(n-l-1)^{-(1+\beta)}\,(1+\bar{\bar{o}}(1)),\quad & p>2+2\,\beta =  \frac{2 D}{2 D-3}
\end{array}
\right.
\end{equation}
(*)Cosine-Bessel regime\\
where the constants $C, C_{B}, C_{A}$ are given by
\begin{equation}\label{2.6}
C(\beta,p):=\displaystyle\frac{2^{\beta+1}}{\pi^{p+1/2}}\,\displaystyle\frac{\Gamma(\beta+1-p/2)\,\Gamma(1-p/2)\,\Gamma(p+1/2)}
{\Gamma(\beta+2-p)\,\Gamma(1+p)}\;,
\end{equation}
\begin{equation}\label{2.5}
C_{A}(p):=
\int_{-\infty}^{+\infty}
\left[
\frac{2\pi}{\sqrt[3]{2}}\,\, {\rm Ai}^2\left(
-\frac{t \sqrt[3]{2}}{2}
\right) \right]^p dt\,,
\end{equation} 
and 
\begin{equation}\label{2.4}
C_{B}(\alpha,\beta,p):=2\int\limits_{0}^{\infty}t^{2\beta+1}|J_{\alpha}|^{2p}(2t)\,dt\;,
\end{equation}
respectively, $\alpha=2l+D-2$ and $\beta=(2-D)p+D-1$.
The symbols $Ai(t)$ and $J_{\alpha}(z)$ denote the Airy and the Bessel functions (see \cite{szego_75}) defined by 
$$
Ai(y)=\frac{\sqrt[3]{3}}{\pi}\,A(-3\sqrt{3}y),\quad A(t)=\frac{\pi}{3}\,\sqrt{\frac{t}{3}}\, \left[J_{-1/3}\left(2\left(\frac{t}{3}\right)^{\frac{3}{2}}\right)+J_{1/3}\left(2\left(\frac{t}{3}\right)^{\frac{3}{2}}\right)\right].
$$
and
$$
J_{\alpha}(z)=\sum_{\nu=0}^{\infty}\frac{(-1)^{\nu}}{\nu!\,\Gamma(\nu+\alpha+1)}\,\left(\frac{z}{2}\right)^{\alpha+2\nu}\;.
$$
respectively.

\textit{Comment}: For $D=3$ it happens that $\frac{2D}{2D-3} =\frac{D-1}{D-2} = 2$, and then the quantities $N_{n,l}(D,p)$ are given by the third asymptotical expression in (\ref{8}). For higher dimensions one has $\frac{2D}{2D-3} >\frac{D-1}{D-2}$, and the three expressions in (\ref{8}) hold.
\end{enumerate}

\textit{Hints}. We use the effective Aptekarev et al's technique \cite{aptekarev_2012,aptekarev_2010,aptekarev_2016} recently applied for oscillator-like systems. This technique determines the ($n\to\infty$)-asymptotics of the Laguerre hydrogenic norms $N_{n}(\alpha,p,\beta)$  by taking care of the values of the parameters $\alpha,\beta$ and $p$. It turns out that the dominant contribution to the asymptotical value of the integral (\ref{eq:c1.2}) comes from different regions of integration defined according to the values of the involved parameters, which characterize various asymptotic regimes. Thus, we have to use various asymptotical representations for the Laguerre polynomials at the different scales.

Altogether there are five asymptotical regimes which can give (depending on $\alpha,\beta$ and $p$) the dominant contribution in the
asymptotics of $N_{n}(\alpha, p, \beta)$. Three of them exhibit the growth of $N_{n}(\alpha, p, \beta)$ as some $n$th-power law with
an exponent which depends on $\alpha,\beta$ and $p$. We call them by Bessel, Airy and cosine (or oscillatory) regimes, which are characterized by the constants $C_{B}$, $C_{A}$ and $C$, respectively, mentioned above. The Bessel regime corresponds to the neighborhood of zero (i.e. the left end point of the interval of orthogonality), where the Laguerre polynomials can asymptotically be expressed by means of Bessel functions (taken for expanding scale of the variable). Then, at the right of zero (cosine regime) the oscillatory behavior of the polynomials (in the bulk region of zeros location) is modeled asymptotically by means of the trigonometric functions; and at the neighborhood of the extreme right of zeros (Airy regime) the asymptotics of the Laguerre polynomials is controlled by Airy functions. Finally, in the neighborhood of the infinity point of the orthogonality interval the polynomials have  growing asymptotics. Moreover, there are regions where these asymptotics match each other. Namely, asymptotics of the Bessel functions for big arguments match the trigonometric function, as well as the asymptotics of the Airy functions do the same. 

In addition, there are two transition regimes: cosine-Bessel and cosine-Airy. If the contributions of these regimes dominate in the integral (\ref{eq:c1.2}), then the asymptotics of $N_{n}(\alpha, p, \beta)$ besides the degree on $n$ have the factor $\ln n$. Note also that if these regimes dominate, then the gamma factors in constant $C(\beta,p)$ in (\ref{2.6}) for the oscillatory cosine regime
blow up. For the cosine-Bessel regime this happens when $\beta+1-p/2=0$, and for the cosine-Airy regime when $1-p/2=0$.
 
\section{Information entropies of the $D$-dimensional Rydberg states}

In this section we determine the Rényi, Shannon and Tsallis entropies of the $D$-dimensional Rydberg hydrogenic states in terms of the spatial dimension $D$, the order parameter $p$, the hyperquantum numbers $(n,l,\{\mu\})$ and the nuclear charge $Z$. First, attention is focussed on the Rényi and Shannon entropies since the Tsallis entropy can be obtained from the Rényi one by means of the relation
\begin{equation}
\label{eq:tsalren}
T_{p}[\rho] = \frac{1}{1-p}[e^{(1-p)R_{p}[\rho]}-1].
\end{equation}
Then, for illustration, we numerically discuss the Rényi entropy $R_{p}[\rho_{n,0,0}]$ of some hydrogenic Rydberg $(ns)$-states in terms of $D$, $p$, $n$ and $Z$.\\
 
Let us start by pointing out that the total Rényi entropy $R_{p}[\rho_{n,l,\{\mu\}}]$ of the Rydberg states can be obtained in a straightforward manner by taking into account Eq. (\ref{eq:renyihyd1}), the values of the radial Rényi entropy $R_{p}[\rho_{n,l}]$ derived from Eq. (\ref{eq:renyihyd4}), the asymptotical ($n \to \infty$) values of the Laguerre norms $N_{n,l}(D,p)$ given by the previous theorem, and the angular Rényi entropy $R_{p}[\mathcal{Y}_{l,\{\mu\}}]$ given by Eqs. (\ref{eq:angpart}) and (\ref{eq:renyihyd3}); keep in mind that the angular part of the Rényi entropy does not depend on the principal quantum number $n$.\\

What about the Shannon entropy $S[\rho_{n,l,\{\mu\}}]$ of the Rydberg hydrogenic states?. To calculate its value for any stationary state $(n,l,\{\mu\})$, we take into account (a) that $\lim_{p \to +1} R_p[\rho] = S[\rho]$ for any probability density $\rho$, (b) the expression (\ref{eq:renyihyd1}), (c) the following limiting value of the radial Rényi entropy $R_{p}[\rho_{n,l,\{\mu\}}]$ of the Rydberg hydrogenic states obtained for a fixed dimension $D$ from (\ref{eq:renyihyd4})-(\ref{eq:c1.21}) and the previous theorem,
\begin{eqnarray}
\label{eq:radshan}
\lim_{p \to +1} R_p[\rho_{n,l}] &=& \lim_{p \to +1} \frac{1}{1-p} \ln \left[\frac{\eta^{D(1-p)-p}}{2^{D(1-p)+p}Z^{D(1-p)}} C(\beta,p)\,(2n)^{1+\beta-p}\right]\nonumber\\
&=& 2 D \ln n +(2-D) \ln 2 +\ln\pi -D\ln Z+D-3,
\end{eqnarray} 
(d) the condition $n>>l$ and (e) that
\begin{equation}
\label{eq:angshan}
\lim_{p \to +1} R_{p}[\mathcal{Y}_{l,\{\mu\}}] = \lim_{p \to +1} \frac{1}{1-p} \ln \Lambda_{l,m}(\Omega_{D-1}) = S[\mathcal{Y}_{l,\{\mu\}}],
\end{equation}
(remember (\ref{eq:angpart}) and (\ref{eq:renyihyd1}) for the first equality) where $S[\mathcal{Y}_{l,\{\mu\}}]$ is the Shannon-entropy functional of the spherical harmonics \cite{dehesa2} given by 
\begin{equation}
S[\mathcal{Y}_{l,\{\mu\}}] = -\int_{S^{D-1}}[\mathcal{Y}_{l,\{\mu\}}]^{2}\, \ln \,[\mathcal{Y}_{l,\{\mu\}}]^{2}\,d\Omega_{D-1}.
\end{equation}
which does not depend on $n$ and can be calculated as indicated in \cite{lopez_2009}. In particular, from the previous indications we find the following values
\begin{equation}
\label{eq:nsrenyi}
R_{p}[\rho_{n,0,0}] = R_p[\rho_{n,0}] + R_{p}[\mathcal{Y}_{0,0}] = R_p[\rho_{n,0}] + \frac{1}{1-p}\ln f(p,D)
\end{equation}
for $p \neq 1$, and
\begin{equation}
\label{eq:sn00}
S[\rho_{n,0,0}] = 2 D \ln n +(2-D) \ln 2 +\ln\pi -D\ln Z+D-3 + S(\mathcal{Y}_{0,0}) + o(1)	
\end{equation}
for the Rényi and Shannon entropy of the ($n$\textit{s})-Rydberg hydrogenic states, respectively, where $f(p,D)$ and $S(\mathcal{Y}_{0,0})$ have the values (see Appendix A):
\begin{eqnarray}
f(p,D) &=& \int_{\Omega_{D-1}} |\mathcal{Y}_{0,0}(\Omega_{D-1})|^{2p}\, d\Omega_{D-1}\nonumber \\
&=& 2^{D(1-p)}\pi^{\frac{1}{2} (-Dp+D+p-1)}\left[\frac{\Gamma (D)}{\Gamma \left(\frac{D+1}{2}\right)}\right]^{p-1}
\end{eqnarray}
and 
\begin{eqnarray}
\label{eq:shany00}
S(\mathcal{Y}_{0,0}) &=& -\ln \mathcal{N}_{0,0}^{2}\nonumber \\
&=& D\ln 2 + \frac{D-1}{2}\ln\pi +\ln \frac{\Gamma \left(\frac{D+1}{2}\right)}{\Gamma (D)},\nonumber\\
\label{eq:harmorenyi}
\end{eqnarray}
respectively. Note that $\lim_{p \to +1} \frac{1}{1-p}\ln f(D,p) = S(\mathcal{Y}_{0,0})$. In the particular case $D=3$ (i.e., for real hydrogenic systems), one has that $\lim_{p \to +1} \frac{1}{1-p}\ln f(3,p) = S(Y_{0,0}) = \ln (4\pi)$, as expected.\\

\begin{figure}[H]
   \centering
       \includegraphics[width=0.5\textwidth]{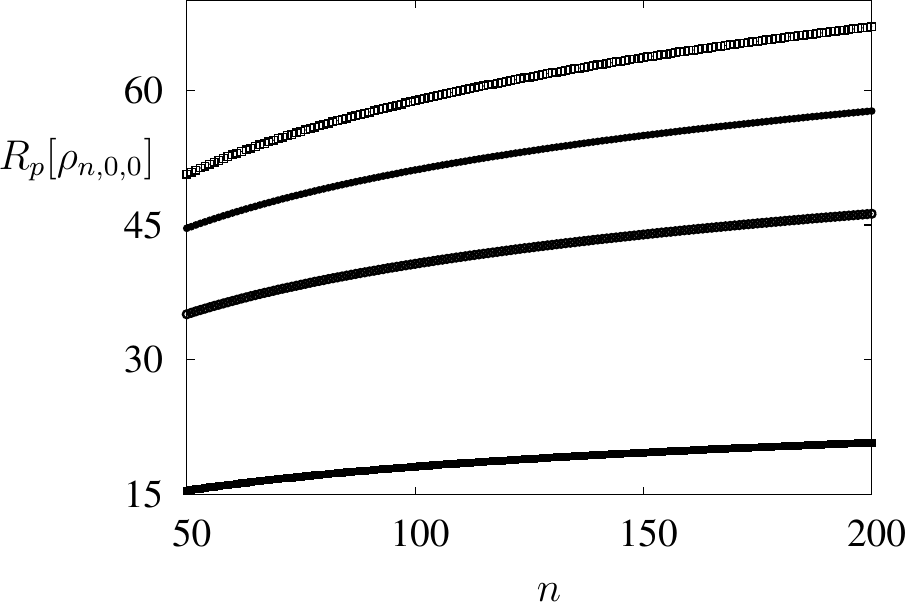}
   \caption{Variation of the Rényi entropy, $R_{p}[\rho]$, with respect to $n$ for the Rydberg hydrogenic $(ns)$-states with $(p=\frac{5}{4}, D=6)(\square)$, $(p=\frac{10}{7}, D=5)(\bullet))$, $(p=\frac{3}{4}, D=4)(\circ))$ and $(p=3, D=2)(\blacksquare))$. }
   \label{fig.1}
 \end{figure}

\textcolor{red}{To conclude and} for illustrative purposes, we first numerically investigate the dependence of the Rényi entropy $R_{p}[\rho_{n,0,0}]$ for some Rydberg $(ns)$-states on the quantum number $n$, the order parameter $p$ and the nuclear charge $Z$. \textcolor{red}{To start with}, we study the variation of the $p$-th order Rényi entropy of these states in terms of $n$, within the interval $n= 50-200$, when $p$ is fixed. As an example, the cases $p=\frac{5}{4},\frac{10}{7},\frac{3}{4},3$ for the $D$-dimensional hydrogenic Rydberg $(ns)$-states with $D=6,5,4$, and $2$, respectively, are plotted in Fig. \ref{fig.1}. We observe that the behavior of the Rényi entropy has always an increasing character for any dimensionality $D>2$. \\

\noindent

\begin{figure}[H]
      \centering
          \includegraphics[width=0.5\textwidth]{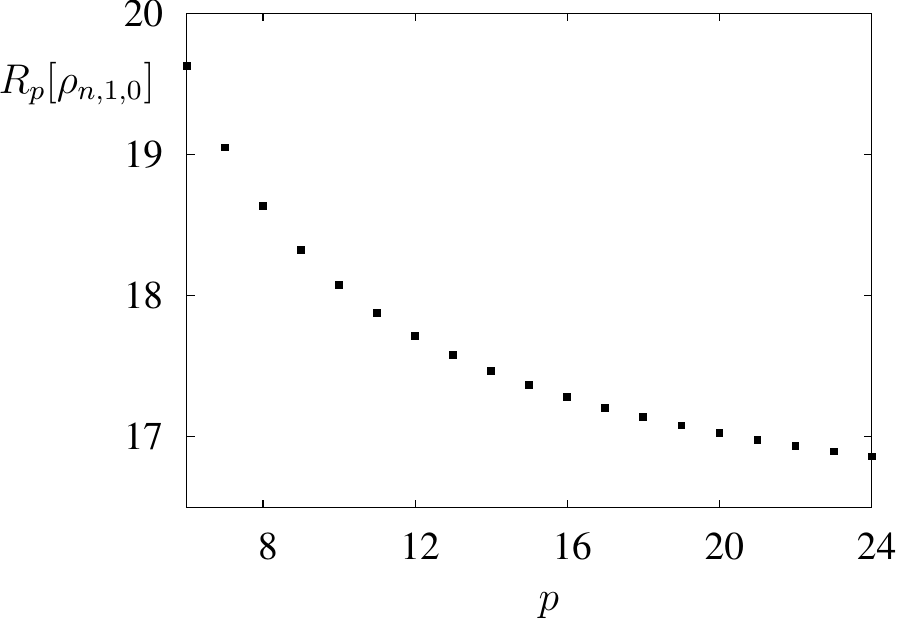}
      \caption{Variation of the Rényi entropy $R_{p}[\rho_{n,1,0}]$, respectively, with respect to $p$ for the Rydberg state $(n=100,l=1)$ of the hydrogen atom ($Z=1$) with $D=4$.}
      \label{fig.2}
    \end{figure}

\textcolor{red}{Then}, in Fig. \ref{fig.2}, we analyze the dependence of the Rényi entropy, $R_{p}[\rho]$, on the order $p$ for the Rydberg hydrogenic state $(n=100, l=1, D=4)$. We observe that the Rényi entropy decreases monotonically as the integer order $p$ is increasing. This behavior indicates that the quantities with the lowest orders (particularly the cases $p=1$ and $p=2$, which correspond to the Shannon entropy and the disequilibrium or second-order Rényi entropy) are most significant for the quantification of the spreading of the electron distribution of the system. In fact, this behavior occurs for all the $D$-dimensional states; we should expect it since the Rényi entropy is defined by (\ref{eq:renentrop}) as a continuous and non-increasing function in $p$. 

\begin{figure}[H]
      \centering
          \includegraphics[width=0.5\textwidth]{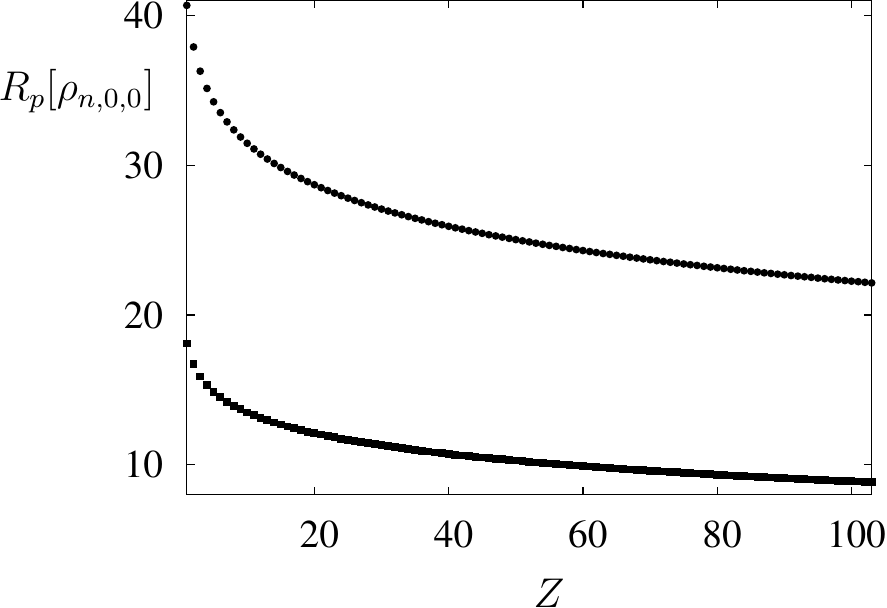}
      \caption{Variation of the Rényi entropy, $R_{p}[\rho]$, with respect to the nuclear charge $Z$ for the Rydberg hydrogenic state $(n=100, l=0)$ of the $D$-dimensional hydrogen atom with $D=2(\blacksquare)$ and $D=4(\bullet)$.}
      \label{fig.3}
    \end{figure}

\noindent
\textcolor{red}{Later}, in Fig. \ref{fig.3}, we illustrate the behavior of the Rényi entropy, $R_{p}[\rho]$, as a function of the atomic number $Z$ of the Rydberg hydrogenic states $(n=100,l=0)$ with $(p=3, D=2)$ and $(p=\frac{3}{4},D=4)$, where $Z$ goes from $1$ (hydrogen) to $103$ (lawrencium). We observe that the Rényi entropy decreases monotonically as $Z$ increases, which points out the fact that the probability distribution of the system tends to separate out from equiprobability more and more as the number of electrons in the nucleus of the atom increases; so, quantifying the greater complexity of the system as the atomic number grows.\\

\begin{figure}[H]
      \centering
          \includegraphics[width=0.5\textwidth]{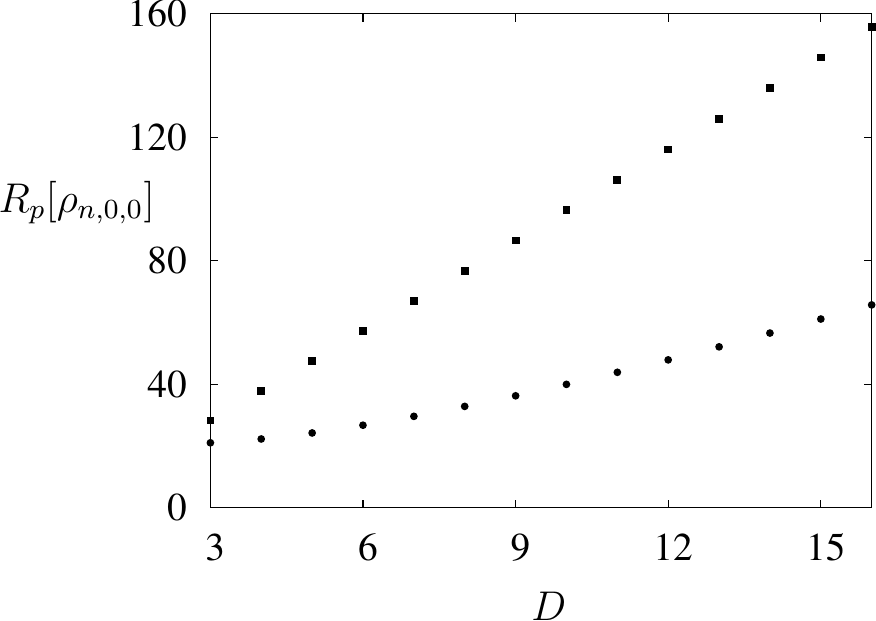}
      \caption{Variation of the Rényi entropy, $R_{p}[\rho]$, with respect to the dimensionality $D$ for the Rydberg hydrogenic state $(n=100, l=0)$ of the $D$-dimensional hydrogen atom with $p=\frac{1}{2}(\blacksquare)$ and $p=4(\bullet)$.}
      \label{fig.4}
    \end{figure}

\noindent
Finally, we investigate the behavior of the Rényi entropy, $R_{p}[\rho]$, of the Rydberg hydrogenic state as a function of the dimensionality $D$. We show it in Fig. \ref{fig.4} for the Rydberg state $(n=100, l=0)$ with $p=\frac{1}{2}$ and $4$ of the hydrogen atom with various integer values of the dimensionality $D \in [50,200]$. We observe that in both cases the Rényi entropy has a monotonically increasing behavior as $D$ grows, which indicates that the larger the dimension, more classically the system behaves (or in other words, the closer is the system to its classical counterpart).

\section{Conclusions}

In this work we determine the three main entropic measures (Rényi, Shannon, Tsallis) of the quantum probability density of stationary $D$-dimensional Rydberg ($n >> 1$) hydrogenic states in terms of the basic parameters which characterize them; namely, the dimensionality $D$, the hyperquantum numbers $(n,l,\{\mu\})$ and the nuclear charge $Z$ of the system. In fact, the Shannon and Tsallis entropies can be obtained from the Rényi one. The Rényi entropy has been calculated by first decomposing it into two parts of radial and angular types, and realizing that the angular part does not depend on $n$, so that the true problem to be solved is the calculation of the radial Rényi entropy in the limit of large $n$. The radial Rényi entropy has been shown to be expressed in terms of the $\mathfrak{L}_{p}$-norms of the Laguerre polynomials which control the Rydberg states we are interested in. Then, the remaining asymptotics of these Laguerre norms is determined by means of a recent technique of approximation theory. Finally, we numerically apply this theoretical methodology to some particular Rydberg hydrogenic states of $s$ and $p$ types. We find that the Rényi entropy monotonically decreases (increases) when the nuclear charge (the dimensionality) is decreasing (increasing) for some $s$ and $p$ states.
\section*{Acknowledgments}
This work has been partially supported by the Projects
FQM-7276 and FQM-207 of the Junta de Andaluc\'ia and the MINECO-FEDER grants
FIS2014-54497P and FIS2014-59311-P. I. V. Toranzo acknowledges the support of ME under the program FPU. We acknowledge useful discussions with Alexander I. Aptekarev.

\appendix

\section{Calculation of $f(p,D)$ and $S(\mathcal{Y}_{0,0})$}

Let us first calculate the factor $f(p,D)$ which appears in Eq. (\ref{eq:nsrenyi}):

\begin{eqnarray}
f(p,D) &=& \int_{\Omega_{D-1}} |\mathcal{Y}_{0,0}(\Omega_{D-1})|^{2p}\, d\Omega_{D-1}\nonumber \\
&=& \int_{\Omega_{D-1}} (\mathcal{N}_{0,0}^{2})^{p}\, d\Omega_{D-1}\nonumber \\
&=& (\mathcal{N}_{0,0}^{2})^{p}\,2 \pi  \prod _{j=1}^{D-2} \frac{\sqrt{\pi } \,\Gamma \left(\frac{D-j}{2}\right)}{\Gamma \left(\frac{1}{2} (D-j+1)\right)}\nonumber \\
&=& (2 \pi )^{1- p}\left[\prod _{j=1}^{D-2} \frac{(D-j-1) \Gamma \left(\frac{1}{2} (D-j-1)\right)^2}{\pi \, 2^{-D+j+3} \Gamma (D-j-1)}\right]^{p}\prod _{j=1}^{D-2} \frac{\sqrt{\pi }\, \Gamma \left(\frac{D-j}{2}\right)}{\Gamma \left(\frac{1}{2} (D-j+1)\right)},\nonumber\\
&=& 2^{1- p}\pi^{\frac{D}{2}(1-p)}\left[\prod _{j=1}^{D-2} \frac{\Gamma \left(\frac{1}{2} (D-j+1)\right)}{\Gamma \left(\frac{D-j}{2}\right)}\right]^{p}\prod _{j=1}^{D-2}\frac{\Gamma \left(\frac{D-j}{2}\right)}{\Gamma \left(\frac{1}{2} (D-j+1)\right)}\nonumber \\
&=&2^{1- p}\pi^{\frac{D}{2}(1-p)} \left[ \Gamma\left(\frac{D}{2}\right) \right]^{p-1}\nonumber\\
\end{eqnarray}

Let us now compute the factor $S(\mathcal{Y}_{0,0})$ which appears in Eq. (\ref{eq:sn00}):
\begin{eqnarray}
\label{eq:shany00}
S(\mathcal{Y}_{0,0}) &=& -\ln \mathcal{N}_{0,0}^{2}\nonumber \\
&=&  -\ln\left[\frac{1}{2 \pi }\prod _{j=1}^{D-2} \frac{(D-j-1) \Gamma \left(\frac{1}{2} (D-j-1)\right)^2}{\pi\,  2^{-D+j+3}\Gamma (D-j-1)}\right],\nonumber \\
&=& \ln 2 \pi - \ln \prod _{j=1}^{D-2} \frac{\Gamma \left(\frac{1}{2} (D-j+1)\right)}{\sqrt{\pi } \,\Gamma \left(\frac{D-j}{2}\right)}\nonumber \\
&=& \ln 2 \pi - \left(1-\frac{D}{2}\right)\ln \pi  - \ln\prod_{j=1}^{D-2}\frac{\Gamma \left(\frac{1}{2} (D-j+1)\right)}{\Gamma \left(\frac{D-j}{2}\right)}\nonumber\\ 
&=& \ln 2 \pi - \left(1-\frac{D}{2}\right)\ln \pi  - \ln \Gamma\left( \frac{D}{2} \right)\nonumber \\
&=& \ln 2 +\frac{D}{2}\ln \pi - \ln \frac{2^{1-D}\pi^{1/2}(D-1)!}{\left(\frac{D-1}{2}\right)!} \nonumber\\
&=& D\ln 2 + \frac{D-1}{2}\ln\pi +\ln \frac{\Gamma \left(\frac{D+1}{2}\right)}{\Gamma (D)}.\nonumber\\
\label{eq:harmorenyi}
\end{eqnarray}

\end{document}